\documentclass[twocolumn,prl,letterpaper,groupedaddress]{revtex4}

\usepackage{times,amsmath,amsfonts,amssymb,latexsym,textcomp}
\usepackage{graphicx}
\usepackage{epstopdf}
\usepackage{color}
\usepackage{epsf}
\usepackage{amsmath}
\usepackage{dsfont}
\usepackage{todonotes}
\usepackage{here}

\renewcommand{\phi}{\varphi}

\renewcommand{\epsilon}{\varepsilon}

\def\ket#1{{\lvert}#1\rangle}
\def\bra#1{\langle #1{\lvert}}
\def\ev#1#2{{\bra{#2}}#1{{\ket{#2}}}}

\input{QCircuit}
\begin{document}
\title{Solving systems of linear equations on a quantum computer}
\author{
\vspace{.2cm}
Stefanie Barz$^1$, Ivan Kassal$^{2,3}$, Martin Ringbauer$^{1,*}$, Yannick~Ole~Lipp$^1$, Borivoje Daki{\'c}$^1$, \\Al\'an~Aspuru-Guzik$^2$, Philip Walther$^1$}
\affiliation{
\vspace{.2cm}
$^1$~Faculty of Physics, University of Vienna, Boltzmanngasse 5, 1090 Vienna, Austria\\
$^2$~Department of Chemistry and Chemical Biology, Harvard University, Cambridge MA 02138, United States\\
$^3$~Centre for Engineered Quantum Systems, Centre for Quantum Computing and Communication Technology, and School of Mathematics and Physics, The University of Queensland, St Lucia QLD 4072, Australia\\
$^*$~Present address: Centre for Engineered Quantum Systems, Centre for Quantum Computing and Communication Technology, and School of Mathematics and Physics, The University of Queensland, St Lucia QLD 4072, Australia}

\begin{abstract}
\vspace{.6cm}
Systems of linear equations are used to model a wide array of problems in all fields of science and engineering.
Recently, it has been shown that quantum computers could solve linear systems exponentially faster than
classical computers~\cite{Harrow2009}, making for one of the most promising applications of quantum computation~\cite{Childs2009}.
Here, we demonstrate this quantum algorithm by implementing various instances on a photonic quantum computing architecture.
Our implementation involves the application of two consecutive entangling gates on the same pair of polarisation-encoded qubits.
We realize two separate controlled-NOT gates where the successful operation of the first gate is heralded by a measurement of two ancillary photons.
Our work thus demonstrates the implementation of a quantum algorithm with high practical significance as well as an important technological advance which brings us closer to a comprehensive control of photonic quantum information~\cite{Knill2001}.
\vspace{1.0cm}
\end{abstract}

\maketitle

%%%%%%%%%%%%%%%%%%%%%%%%%%%%%%%%%%%%%%%%%%%%%%%%%%%%%%%%%%%%%%%%%%%%%%%%%%%%%%%%%%%%%%%%%%%%%%%%%%%%%%%%%%%%%%%%%%%%%%%%%%%%%%%%%%%%%%%%%%%%%%%%%%%%%%
%
%\section{Introduction}
%%%%%%%%%%%%%%%%%%%%%%%%%%%%%%%%%%%%%%%%%%%%%%%%%%%%%%%%%%%%%%%%%%%%%%%%%%%%%%%%%%%%%%%%%%%%%%%%%%%%%%%%%%%%%%%%%%%%%%%%%%%%%%%%%%%%%%%%%%%%%%%%%%%%%%
Systems of linear equations play an important role in various fields, ranging from natural science and engineering to medicine and social science.
The ability to solve such systems underpins many modern technologies, including traffic flow analysis, computer tomography, and weather forecasting.
Although systems of linear equations are an old problem---the two-dimensional version was investigated by the ancient Babylonians---it was only in 1811 that Gauss developed a general algorithm for solving them, Gaussian elimination~\cite{Kleiner2007}. Today, as the sizes of data sets are growing, Gaussian elimination is too slow and more advanced methods are needed.
An increasing demand for a detailed understanding of larger and larger systems pushes the tools of classical computation to their limits.
Modern data sets can be so enormous that finding a solution to a system of linear equations can be prohibitive even for the latest supercomputers. 

Quantum computers have attracted tremendous interest because they could outperform classical computers at certain tasks~\cite{Nielsen2000}.
For a classical computer, the number of computational steps needed to solve a linear system is at least proportional to the number of variables. By contrast, the new quantum algorithm could, in some cases, make the computational time proportional to only the logarithm of the number of variables~\cite{Harrow2009}. 
An important difference is that the quantum algorithm calculates the expectation value of an operator associated with the solution rather than the solution itself.
Here, we demonstrate this algorithm using two full controlled-NOT (CNOT) gates acting on two qubits to determine the solution of a two-dimensional system of linear equations.
We show various instances of the algorithm for systems of linear equations with different characteristics.

%%%%%%%%%%%%%%%%%%%%%%%%%%%%%%%%%%%%%%%%%%%%%%%%%%%%%%%%%%%%%%%%%%%%%%%%%%%%%%%%%%%%%%%%%%%%%%%%%%%%%%%%%%%%%%%%%%%%%%%%%%%%%%%%%%%%%%%%%%%%%%%%%%%%%%
\section{Theory}
%%%%%%%%%%%%%%%%%%%%%%%%%%%%%%%%%%%%%%%%%%%%%%%%%%%%%%%%%%%%%%%%%%%%%%%%%%%%%%%%%%%%%%%%%%%%%%%%%%%%%%%%%%%%%%%%%%%%%%%%%%%%%%%%%%%%%%%%%%%%%%%%%%%%%%
Solving a linear system of equations, given a matrix $A$ and a vector $b$, means finding the vector $x$ such that $Ax=b$.
If we rescale the vectors to $\left\Vert b\right\Vert =\left\Vert x\right\Vert =1$, we can represent them as quantum states $\ket{b}$ and $\ket{x}$,
and our initial task becomes finding the state $\ket{x}$ such that
\begin{equation}
A\ket{x}=\ket{b}.
\end{equation}
The solution we seek is
\begin{equation}
\left|x\right\rangle =\frac{A^{-1}\left|b\right\rangle}{\left\Vert A^{-1}\left|b\right\rangle \right\Vert}.
\end{equation}

%---------------------------------------------------------------------------------------------------------------
\begin{figure*}%[th]
\includegraphics[width =1.25\columnwidth]{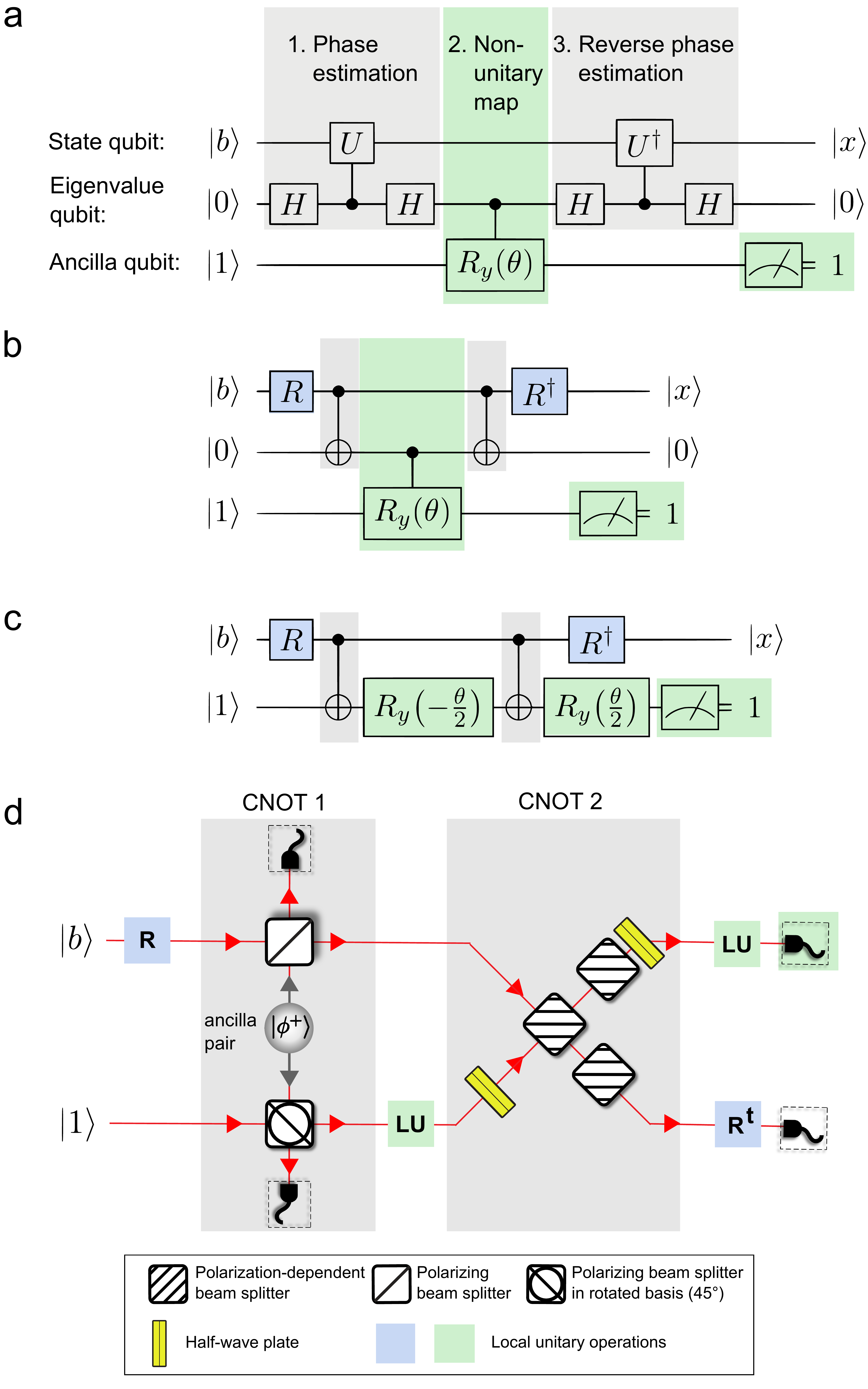}
\caption[~]{The simplest case of the quantum algorithm for solving systems of linear equations. Given a Hermitian
matrix $A$ and input $\ket{b}$, outputs $\left|x\right\rangle =A^{-1}\left|b\right\rangle / \left\Vert A^{-1}\left|b\right\rangle \right\Vert $
if the ancilla qubit is measured to be 1.
\textbf{(a)} The complete circuit, as derived in the text, with $U=\exp\left(2\pi i \,2^{n-1} A\right)$ and $\theta=-2\arccos(\lambda_1/\lambda_2)$, 
where $\lambda_{1,2}$ are the eigenvalues of $A$ and the integer $n$ depends on the eigenvalues (see Appendix). 
\textbf{(b)} The local unitary $R$ diagonalises $A$,
$A =R^{\dagger} \left( \begin{smallmatrix} \lambda_1 & 0\\0&\lambda_2 \end{smallmatrix} \right) R$. 
For the algorithm to work perfectly with this many qubits, $\lambda_{1,2}$ must be such that 
$U=R^{\dagger} \left( \begin{smallmatrix} 1 & 0\\0&-1 \end{smallmatrix} \right)R=R^{\dagger}ZR$.
This circuit reflects this simplification.
\textbf{(c)} Optimised circuit. The middle qubit can be completely removed, and the controlled-rotation decomposed, 
to give the final circuit (see Appendix for details). $\ket{b}$ and $R$ are arbitrary. 
\textbf{(d)} Experimental implementation of the circuit shown in c. The local unitary operations are implemented with the help of a combination of half-wave plates and quarter-wave plates. 
\label{Figure1}}
\end{figure*}
%---------------------------------------------------------------------------------------------------------------

We assume, without loss of generality~\cite{Harrow2009}, that $A$ is an \mbox{$N\times N$} Hermitian matrix with eigenbasis $\left\{ \ket{u_j} \right\} $
and eigenvalues $\left\{ \lambda_{j} \right\}$, and is rescaled so that \mbox{$0< \lambda_{j} <1$}.
The state $\ket{b}$ can be expanded in the eigenbasis, $\left|b\right\rangle =\sum_{j=1}^{N}\beta_{j}\left|u_{j}\right\rangle$,
and we aim to prepare, up to normalization,
\begin{equation}
\left|x\right\rangle =\sum_{j=1}^{N}\beta_{j}\frac{1}{\lambda_{j}}\left|u_{j}\right\rangle .
\end{equation}

The algorithm consists of three main steps. Here, we describe the basic idea of the algorithm; a more detailed discussion can be found in~\cite{Harrow2009} and the Appendix).

The first step is to apply phase estimation, a general procedure for decomposing quantum states in a particular basis~\cite{Kitaev1997, Berry2007,Childs2010}. %~\cite{Kitaev1995,Abrams1999}.
%Using the unitary $U=e^{2\pi iA}$, 
For this, we add an additional ``eigenvalue register'' of $m$ qubits to our system, each initialized in the state $\ket{0}$.
Phase estimation then transforms $\ket{b}\ket{0}^{\otimes m}$ into
$\sum_{j=1}^{N}\beta_{j}\ket{u_{j}}\ket{\lambda_{j}}$, where the eigenvalues $\ket{\lambda_{j}}$ are stored in the eigenvalue register
to a precision of $m$ binary digits.

%---------------------------------------------------------------------------------------------------------------
\begin{figure*}[tb]
\includegraphics[width = 1.5\columnwidth]{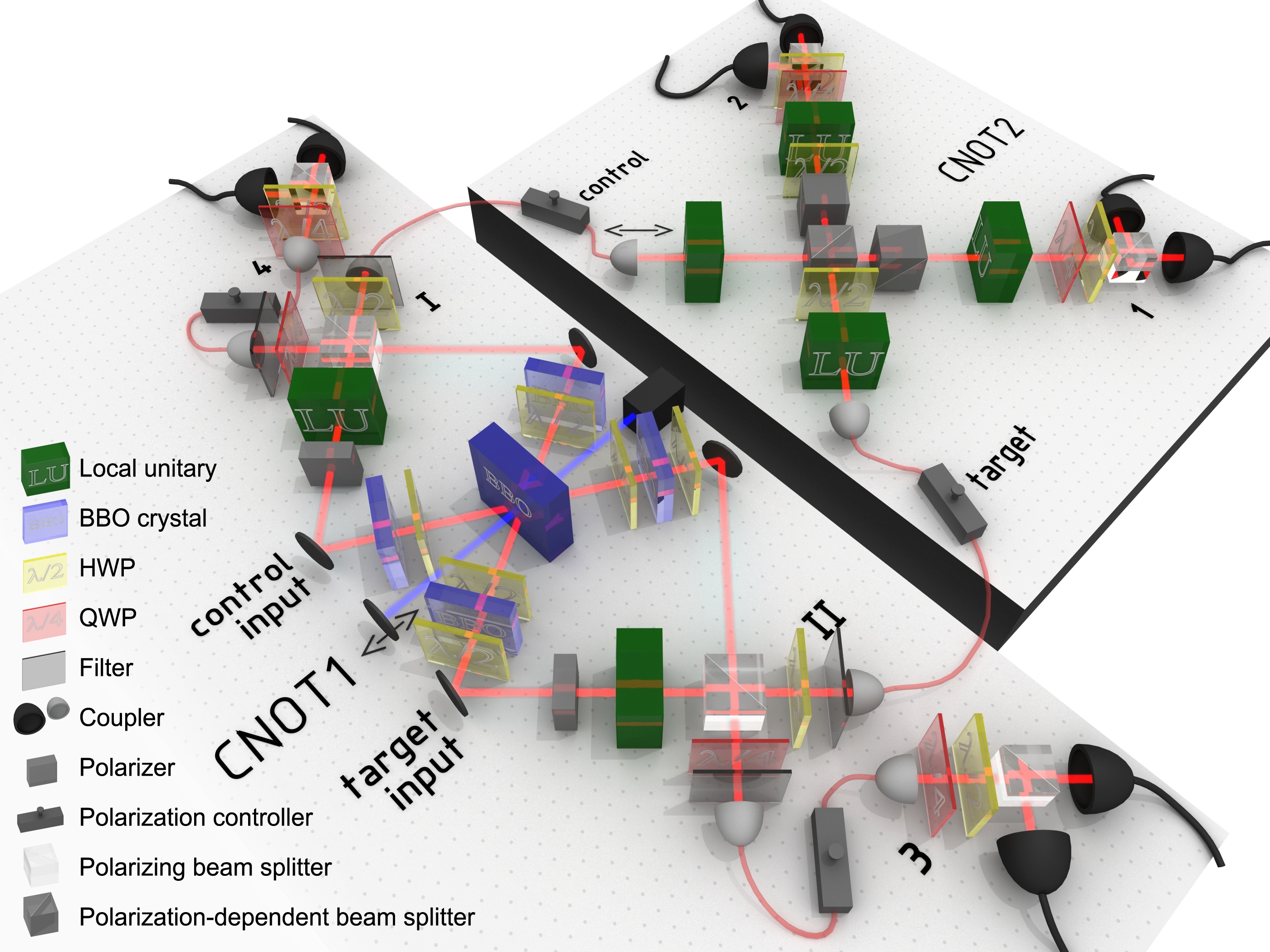}
\caption[~]{Experimental setup. Shown is the experimental implementation of two concatenated CNOT gates. The input is set by a polariser, which can be followed by a local unitary operation (LU). The two gates are connected by fibers. Different matrices $A$ can be implemented by adapting the LUs, and different states $\ket{b}$ by adapting the input state. The figure shows the most general case of two concatenated CNOT gates, combined with general LUs. For the implementation of the algorithm, we chose some LUs to be the identity and obtained the case shown in Figure~\ref{Figure1}d.
\label{Figure2}}
\end{figure*}
%---------------------------------------------------------------------------------------------------------------

The second step is to implement the nonunitary map $\ket{\lambda_{j}}\rightarrow\lambda_{j}^{-1}\ket{\lambda_{j}}$. 
For this, we introduce an additional ``ancilla qubit'' initially in the state $\ket{1}$. Depending on the value of $|\lambda_{j}\rangle$ 
in the eigenvalue register, we implement a controlled $R_y\!\left(\theta_j\right)$ rotation on the ancilla qubit.
Here, $R_y(\theta) = \exp(-i\theta \sigma_y /2)$ and $\sigma_y$ denotes the usual Pauli matrix.
With $\theta_j=-2\arccos(C/\lambda_j)$, the state of our system becomes
\begin{equation}
\sum_{j=1}^{N}\beta_{j}\left|u_{j}\right\rangle \left|\lambda_{j}\right\rangle 
\left(\sqrt{1-\frac{C^{2}}{\lambda_{j}^{2}}}\left|0\right\rangle +\frac{C}{\lambda_{j}}\left|1\right\rangle \right),
\end{equation}
where $C\le\min_{j}|\lambda_{j}|$. 

Finally, the third step is to run phase estimation in reverse to uncompute $\left|\lambda_{j}\right\rangle $, giving
\begin{equation}
\sum_{j=1}^{N}\beta_{j}\left|u_{j}\right\rangle \left(\sqrt{1-\frac{C^{2}}{\lambda_{j}^{2}}}\left|0\right\rangle +\frac{C}{\lambda_{j}}\left|1\right\rangle \right).
\end{equation}
We measure the ancilla qubit, and if we observe a 1 we will have prepared $\left|x\right\rangle $ in the state register. 
If we know the eigenvalues, we can maximise the success probability by choosing the largest possible $C$, $C=\min_j |\lambda_j|$.

The runtime of the algorithm is $\tilde{O}(\log(N) s^2\kappa^2/\epsilon)$,
where $s$ is the sparsity of the matrix, $\kappa$ its
condition number, and $\epsilon$ the acceptable error~\cite{Harrow2009}.
Furthermore, $\tilde{O}(\kappa^2)$ can be reduced to $\tilde{O}(\kappa)$ with amplitude amplification~\cite{Ambainis2010}. 
The best classical algorithms require $O(Ns\sqrt{\kappa}\log(1/\epsilon))$ time,
meaning that at constant $s$, $\kappa$, and $\epsilon$, the quantum
algorithm is exponentially faster. 
On a quantum computer, determining all amplitudes in  
a quantum state scales exponentially with system size; instead, the strength of the algorithm lies
in the determination of expectation values $\langle x|\hat{M}|x\rangle $ of some operator~$\hat{M}$.

The simplest case involves three qubits: one
state qubit, one eigenvalue qubit, and the ancilla
qubit. The state and eigenvalue qubits replace larger
registers which would be needed in a general implementation of the algorithm. 
Whereas using one state qubit means that $\ket{b}$ is a two-vector and $A$ a $2\times2$ matrix,
using one eigenvalue qubit imposes more subtle restrictions.
With one eigenvalue qubit, only a single binary digit of the eigenvalues is computed by the phase estimation,
meaning that for the algorithm to work perfectly, it must be possible to distinguish the two eigenvalues with
a single digit. Consequently, we choose the two eigenvalues to be of the 
form $0.\bar{a}0$ and $0.\bar{a}1$, where $\bar{a}$ is a sequence of binary digits.

The complete circuit for the algorithm as described is given in Figure~\ref{Figure1}, which also outlines the procedure for optimising the circuit to require only two qubits and two consecutive CNOTs acting on them (Figure~\ref{Figure1}c). 
The circuit depends on the eigenvalues of $A$, the unitary $R$ that diagonalizes it,
\begin{equation}
	A= R^\dagger
\begin{pmatrix}
	\lambda_1&0\\
	0        &\lambda_2
\end{pmatrix}R,
\end{equation}
and the input state $\ket{b}$. The algorithm succeeds---i.e. the ancilla qubit is measured in the state $\ket{1}$---with probability $(\lambda_1/\lambda_2)^2$.

In our implementation, we choose different sets of eigenvalues $\Lambda=\left\{\lambda_1, \lambda_2\right\}$.
The simplest case is $\bar{a}=\bar{1}$, giving the eigenvalues $\lambda_{1}=0.10=\frac12$ and $\lambda_{2}=0.11=\frac34$.
In this case phase estimation must read out the second bit of the eigenvalues.
The procedure is analogous for the other sets of eigenvalues that
we implemented (see Appendix).
%%%%%%%%%%%%%%%%%%%%%%%%%%%%%%%%%%%%%%%%%%%%%%%%%%%%%%%%%%%%%%%%%%%%%%%%%%%%%%%%%%%%%%%%%%%%%%%%%%%%%%%%%%%%%%%%%%%%%%%%%%%%%%%%%%%%%%%%%%%%%%%%%%%%%%

\section{Experimental realization}

%%%%%%%%%%%%%%%%%%%%%%%%%%%%%%%%%%%%%%%%%%%%%%%%%%%%%%%%%%%%%%%%%%%%%%%%%%%%%%%%%%%%%%%%%%%%%%%%%%%%%%%%%%%%%%%%%%%%%%%%%%%%%%%%%%%%%%%%%%%%%%%%%%%%%%
%---------------------------------------------------------------------------------------------------------------
\begin{figure}%[h]
\includegraphics[width =0.7\columnwidth]{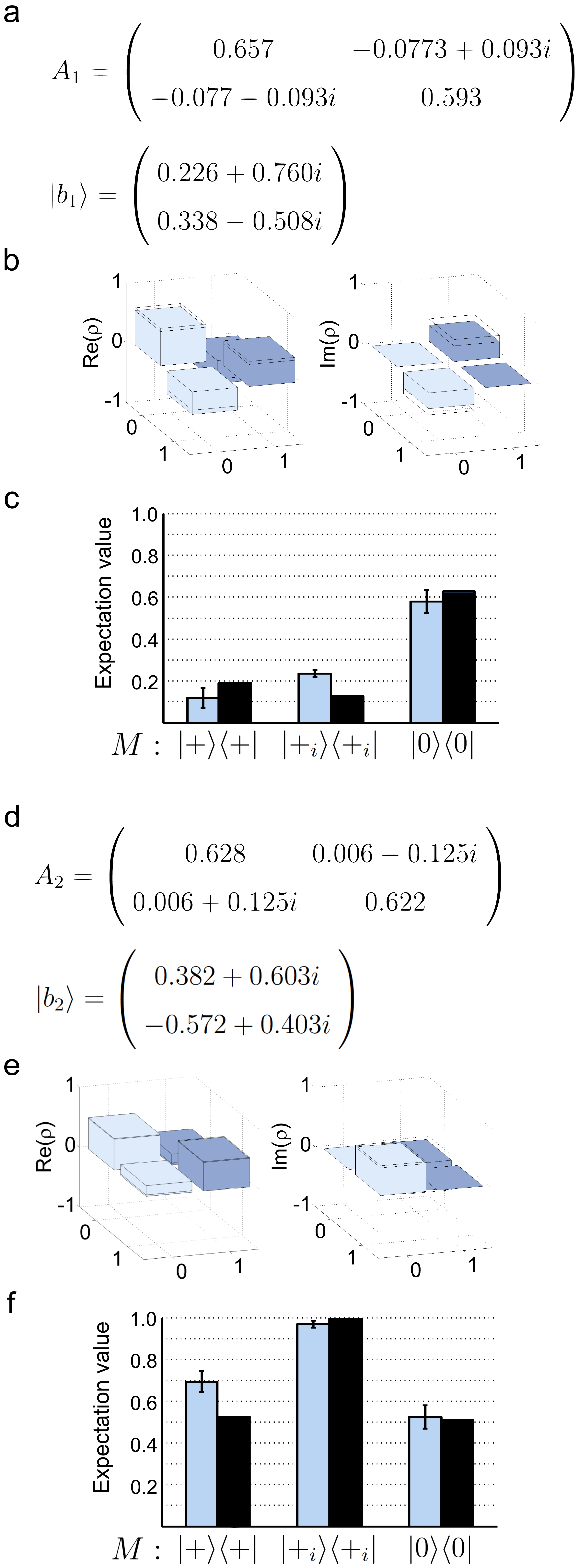}
\caption[~]{Experimental results. 
\textbf{(a), (d)} The figure shows two different systems of linear equations, depicted by the matrices $A_1$ and $A_2$ as well as the state vectors $\ket{b_1}$ and $\ket{b_2}$. 
\textbf{(b), (e)} The reconstructed density matrices of the experimentally obtained output state $\ket{x}$ are shown. These density matrices are obtained by choosing the local operations $R_1=R_x(\frac{11}{15}\pi).R_y(\frac{3}{8}\pi)$ (b), and $R_2=R_x(\frac{89}{60}\pi).R_y(-\frac{3}{8}\pi)$ (e).
For both matrices, we choose the eigenvalues to be $\lambda_1=\frac{1}{2}$ and $\lambda_1=\frac{3}{4}$ by implementing $R_y(\theta)$ as described in the main text. The fidelities of the reconstructed density matrices are $0.953\pm0.026$ (b) and $0.976\pm0.010$ (e). The wireframe shows the theoretical prediction.
\textbf{(c), (f)} The quantum algorithm is based on determining the expectation value $\ev{\hat{M}}{x}$ of some operator $\hat{M}$ with respect to the output state $\ket{x}$.
Therefore, we also show the experimentally determined (blue) and theoretical (black) expectation values of several operators $\hat{M}$. We choose the operator $\hat{M}$ to be the projection on the states
 $\ket{0}$, $\ket{+}$, and $\ket{+_i}$, respectively, with $\ket{+}=(\ket{0}+\ket{1})/\sqrt{2}$, and $\ket{+_i}=(\ket{0}+i\ket{1})/\sqrt{2}$.
\label{Figure3}}
\end{figure}
%---------------------------------------------------------------------------------------------------------------

We implemented the algorithm using polarisation-encoded photonic qubits (Figures~\ref{Figure1}d and~\ref{Figure2}), 
where $\ket{0}$ and $\ket{1}$ denote horizontal and vertical polarisation, respectively~\cite{Kok2007, OBrien2009}.
Our implementation uses two different types of photonic CNOT gates~\cite{OBrien2003,Langford2005, Kiesel2005a, Okamoto2005, Zhou2011a, Crespi2011, Sansoni2010}.

The first CNOT gate realizes the gate operation in a heralded manner~\cite{Pittman2001, Gasparoni2004}. It requires an entangled ancilla photon-pair and a measurement of two ancilla modes to herald that the gate has worked correctly.
If two photons are registered in the ancilla modes, the gate has been successful without the need for a verification of the output state. Since the output photons need not to be measured, the application of a second CNOT gate is possible.

Our second CNOT is implemented in a destructive way, where a coincident measurement of the output photons signals the correct gate operation.
The basic element of this destructive CNOT gate is a polarisation-dependent beam splitter (PDBS) which has a different transmission coefficient $T$ for horizontally polarised light ($T_H=1$) as for vertically polarised light ($T_V=1/3$)~\cite{Kiesel2005a}. 
If two vertically-polarised photons are reflected at this PDBS, they acquire a phase shift of $\pi$.
Two successive PDBSs with the opposite splitting ratios then equalize the output amplitudes.
This setup, in combination with two half-wave plates (HWPs) (see Figure~\ref{Figure1}d) implements a destructive CNOT gate.
The gate has been successful if one photon is measured in each output mode. 

Combining these photonic CNOT gates with local unitary operations allows us to implement the circuit shown in Figure~\ref{Figure1}.
In our setup, we implement these local unitary operations with the help of a combination of quarter-wave plates (QWPs) and HWPs (see Figures~\ref{Figure1}d and~\ref{Figure2}). A detailed description of our experimental setup can be found in the Appendix.

%%%%%%%%%%%%%%%%%%%%%%%%%%%%%%%%%%%%%%%%%%%%%%%%%%%%%%%%%%%%%%%%%%%%%%%%%%%%%%%%%%%%%%%%%%%%%%%%%%%%%%%%%%%%%%%%%%%%%%%%%%%%%%%%%%%%%%%%%%%%%%%%%%%%%%

\section{Implementation of the algorithm}

%%%%%%%%%%%%%%%%%%%%%%%%%%%%%%%%%%%%%%%%%%%%%%%%%%%%%%%%%%%%%%%%%%%%%%%%%%%%%%%%%%%%%%%%%%%%%%%%%%%%%%%%%%%%%%%%%%%%%%%%%%%%%%%%%%%%%%%%%%%%%%%%%%%%%%

We have implemented various instances of the algorithm, where we varied both the matrix $A$ and the state $\ket{b}$.
The matrix can be modified by tuning the local operation $R$ (in which case the eigenvalues stay the same) or by adapting $\theta$, 
which alters the eigenvalues of the matrix. The detection of the ancilla qubit in the state $\ket{1}$ announces a successful
run of the algorithm and the preparation of the output qubit in the state $\ket{x}$.

Figure~\ref{Figure3} shows the results of a sample run of the algorithm. 
Two different matrices $A$ were implemented by using different local operations $R_1$ and $R_2$. The eigenvalues were 
$\Lambda=\left\{\frac{1}{2}, \frac{3}{4}\right\}$ in both cases.
We then choose state vectors $\ket{b_1}$ and $\ket{b_2}$ by setting different input states in our experiment (see Figure~\ref{Figure3}a).
We run the algorithm and analyse the output state $\ket{x}$ via quantum state tomography.
%F_a= 0.953 \pm 0.026, F_b=0.975 \pm 0.009
The density matrices of the resulting output states $\ket{x}$ are shown in Figure~\ref{Figure3}c.
As the algorithm relies on the determination of the expectation values $\ev{\hat{M}}{x}$, 
the figure also shows expectation values with several operators $\hat{M}$ (see Figure~\ref{Figure3}d). 
%$R=\Ident,\sigma_X,\sigma_Y,\sigma_Z,R_x(-\frac\pi 2),R_y(-\frac \pi 4), R_x(\frac{89}{60}\pi).R_y(-\frac{3}{8}\pi), R_x(\frac{11}{15}\pi).R_y(\frac{3}{8}\pi)$

Furthermore, we implemented the algorithm for a series of matrices $A$ with three different sets of eigenvalues $\Lambda_1=\left\{\frac{1}{2}, \frac{3}{4}\right\}$, $\Lambda_2=\left\{\frac{1}{2}, \frac{5}{8}\right\}$, and $\Lambda_3=\left\{\frac{3}{4}, \frac{7}{8}\right\}$ (see Figure~\ref{Figure4}).
As explained in detail in the Appendix, the performance of the algorithm depends on the state $R\ket{b}$ that enters the gate. 
In order to analyse this behaviour, we chose two different input states, $\ket{b_1}=\ket{1}$ and $\ket{b_2}=\ket{+}$ for each set of eigenvalues, while keeping $R$ equal to the identity matrix. The six resulting density matrices are shown in Figure~\ref{Figure4}.
We achieve fidelities of up to $0.981 \pm 0.009$ for $\ket{b_1}$, and $0.832 \pm 0.031$ for $\ket{b_2}$. 
This difference in fidelities arises due to the influence of higher-order emissions, which depends on the state $R\ket{b}$, as discussed in the Appendix. 
Additional data for a set of different input states and the three different choices of eigenvalues is shown in the Appendix.

%---------------------------------------------------------------------------------------------------------------
\begin{figure}%[th]
\includegraphics[width = .95\columnwidth]{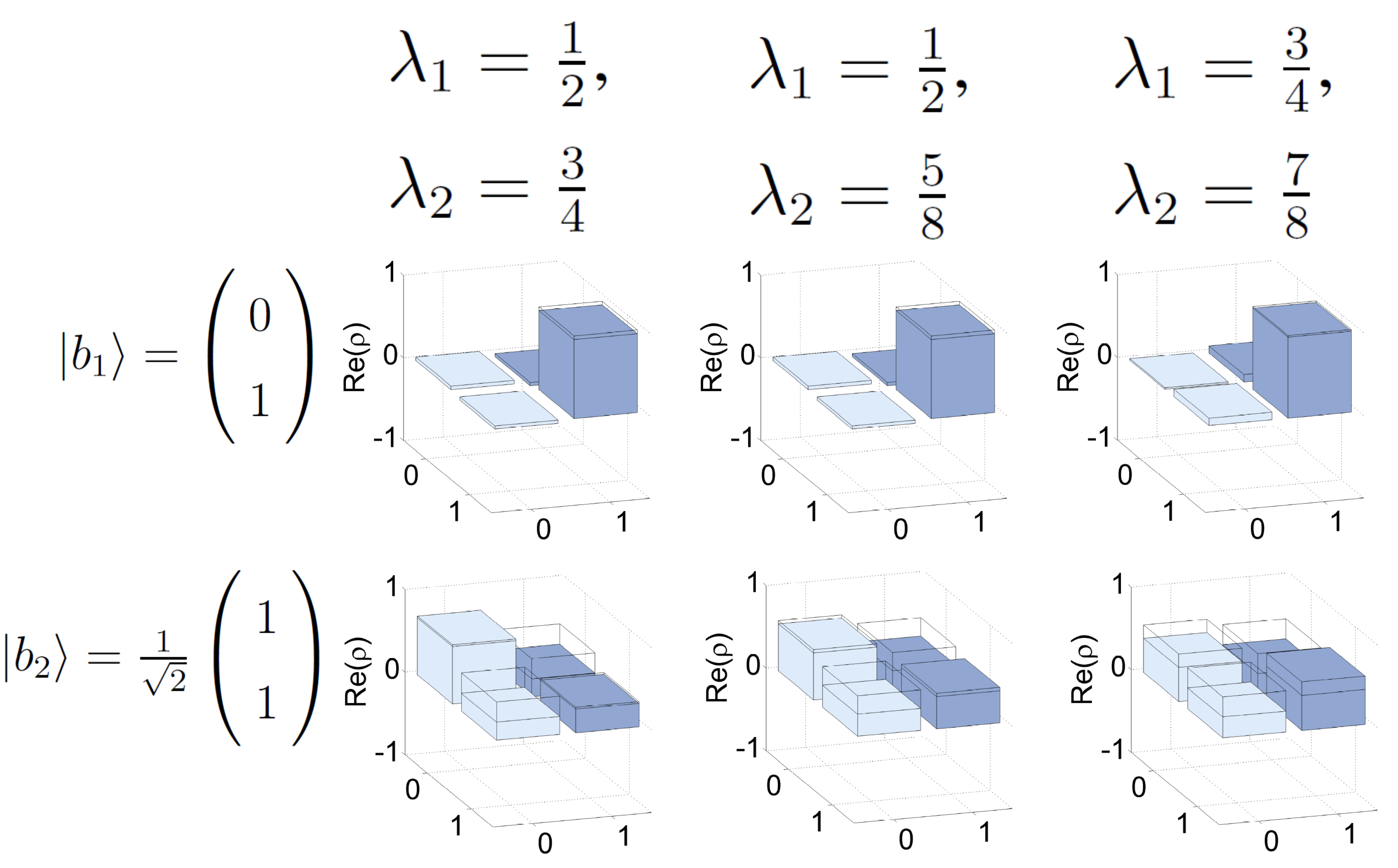}
\caption[~]{The figure shows the solution of the system of linear equation for matrices with different eigenvalues. Experimentally, these are obtained by implementing different values of $\theta$ (see Appendix for details). For all matrices we run the algorithm for two input states $\ket{b_1}=\ket{1}=(0,1)$ and $\ket{b_2}=\ket{+}=(1,1)/\sqrt{2}$. We achieve fidelities of $0.957\pm0.010$, $0.961\pm0.013$, $0.981\pm0.009$ for the input state $\ket{b_1}$ (upper row from left to right) and $0.778\pm0.031$, $0.773\pm0.027$, $0.832\pm0.031$ for the input state $\ket{b_2}$ (lower row, left to right).
\label{Figure4}}
\end{figure}
%---------------------------------------------------------------------------------------------------------------

%%%%%%%%%%%%%%%%%%%%%%%%%%%%%%%%%%%%%%%%%%%%%%%%%%%%%%%%%%%%%%%%%%%%%%%%%%%%%%%%%%%%%%%%%%%%%%%%%%%%%%%%%%%%%%%%%%%%%%%%%%%%%%%%%%%%%%%%%%%%%%%%%%%%%%

\section{Discussion}

%%%%%%%%%%%%%%%%%%%%%%%%%%%%%%%%%%%%%%%%%%%%%%%%%%%%%%%%%%%%%%%%%%%%%%%%%%%%%%%%%%%%%%%%%%%%%%%%%%%%%%%%%%%%%%%%%%%%%%%%%%%%%%%%%%%%%%%%%%%%%%%%%%%%%%

The results presented here include the implementation of the simplest case of the quantum algorithm for solving systems of linear equations, as well as the concatenation of two entangling gates acting on the same photonic qubits.
We anticipate that increasing technological capabilities,
including the implementation of more than two consecutive CNOTs, will allow the extension of the algorithm both to larger systems
and to more precise and full phase estimation.
In particular, the ability to do arithmetic operations with quantum gates will enable the calculation of inverses
(and thus the map $\left|\lambda_{j}\right\rangle \rightarrow\lambda_{j}^{-1}\left|\lambda_{j}\right\rangle $) on the fly.
Even though current quantum computations in optical systems are proof-of-principle demonstrations, important insights can be obtained for future realisations in larger systems~\cite{Lanyon2010,Zhou2011, Martin-Lopez2012}.
We are likewise hopeful that our demonstration of this algorithm will enable future
implementations of other, equally important algorithms that use it as a subroutine, including 
quantum algorithms for solving nonlinear differential equations~\cite{Leyton2008} and quantum data fitting~\cite{Wiebe2012}.

%%%%%%%%%%%%%%%%%%%%%%%%%%%%%%%%%%%%%%%%%%%%%%%%%%%%%%%%%%%%%%%%%%%%%%%%%%%%%%%%%%%%%%%%%%%%%%%%%%%%%%%%%%%%%%%%%%%%%%%%%%%%%%%%%%%%
%%%%%%%%%%%%%%%%%%%%%%%%%%%%%%%%%%%%%%%%%%%%%%%%%%%%%%%%%%%%%%%%%%%%%%%%%%%%%%%%%%%%%%%%%%%%%%%%%%%%%%%%%%%%%%%%%%%%%%%%%%%%%%%%%%%%
%\bibliography{inversion}

%%%%%%%%%%%%%%%%%%%%%%%%%%%%%%%%%%%%%%%%%%%%%%%%%%%%%%%%%%%%%%%%%%%%%%%%%%%%%%%%%%%%%%%%%%%%%%%%%%%%%%%%%%%%%%%%%%%%%%%%%%%%%%%%%%%%
%%%%%%%%%%%%%%%%%%%%%%%%%%%%%%%%%%%%%%%%%%%%%%%%%%%%%%%%%%%%%%%%%%%%%%%%%%%%%%%%%%%%%%%%%%%%%%%%%%%%%%%%%%%%%%%%%%%%%%%%%%%%%%%%%%%%
%\bibliography{inversion}

\begin{thebibliography}{27}
\expandafter\ifx\csname natexlab\endcsname\relax\def\natexlab#1{#1}\fi
\expandafter\ifx\csname bibnamefont\endcsname\relax
  \def\bibnamefont#1{#1}\fi
\expandafter\ifx\csname bibfnamefont\endcsname\relax
  \def\bibfnamefont#1{#1}\fi
\expandafter\ifx\csname citenamefont\endcsname\relax
  \def\citenamefont#1{#1}\fi
\expandafter\ifx\csname url\endcsname\relax
  \def\url#1{\texttt{#1}}\fi
\expandafter\ifx\csname urlprefix\endcsname\relax\def\urlprefix{URL }\fi
\providecommand{\bibinfo}[2]{#2}
\providecommand{\eprint}[2][]{\url{#2}}

\bibitem[{\citenamefont{Harrow et~al.}(2009)\citenamefont{Harrow, Hassidim, and
  Lloyd}}]{Harrow2009}
\bibinfo{author}{\bibfnamefont{A.}~\bibnamefont{Harrow}},
  \bibinfo{author}{\bibfnamefont{A.}~\bibnamefont{Hassidim}}, \bibnamefont{and}
  \bibinfo{author}{\bibfnamefont{S.}~\bibnamefont{Lloyd}},
  \bibinfo{journal}{Phys. Rev. Lett.} \textbf{\bibinfo{volume}{103}},
  \bibinfo{pages}{150502} (\bibinfo{year}{2009}).

\bibitem[{\citenamefont{Childs}(2009)}]{Childs2009}
\bibinfo{author}{\bibfnamefont{A.}~\bibnamefont{Childs}},
  \bibinfo{journal}{Nature Physics} \textbf{\bibinfo{volume}{5}},
  \bibinfo{pages}{861} (\bibinfo{year}{2009}).

\bibitem[{\citenamefont{Knill et~al.}(2001)\citenamefont{Knill, Laflamme, and
  Milburn}}]{Knill2001}
\bibinfo{author}{\bibfnamefont{E.}~\bibnamefont{Knill}},
  \bibinfo{author}{\bibfnamefont{R.}~\bibnamefont{Laflamme}}, \bibnamefont{and}
  \bibinfo{author}{\bibfnamefont{G.~J.} \bibnamefont{Milburn}},
  \bibinfo{journal}{Nature} \textbf{\bibinfo{volume}{409}}, \bibinfo{pages}{46}
  (\bibinfo{year}{2001}).

\bibitem[{\citenamefont{Kleiner}(2007)}]{Kleiner2007}
\bibinfo{author}{\bibfnamefont{I.}~\bibnamefont{Kleiner}},
  \emph{\bibinfo{title}{A history of abstract algebra}}
  (\bibinfo{publisher}{Birkh{\"a}user}, \bibinfo{year}{2007}).

\bibitem[{\citenamefont{Nielsen and Chuang}(2000)}]{Nielsen2000}
\bibinfo{author}{\bibfnamefont{M.~A.} \bibnamefont{Nielsen}} \bibnamefont{and}
  \bibinfo{author}{\bibfnamefont{I.~L.} \bibnamefont{Chuang}},
  \emph{\bibinfo{title}{{Quantum Computation and Quantum Information}}}
  (\bibinfo{publisher}{Cambridge University Press}, \bibinfo{year}{2000}).

\bibitem[{\citenamefont{Kitaev}(1997)}]{Kitaev1997}
\bibinfo{author}{\bibfnamefont{A.}~\bibnamefont{Kitaev}},
  \bibinfo{journal}{Russ. Math. Surv.} \textbf{\bibinfo{volume}{52}},
  \bibinfo{pages}{1191} (\bibinfo{year}{1997}).

\bibitem[{\citenamefont{Berry et~al.}(2007)\citenamefont{Berry, Ahokas, Cleve,
  and Sanders}}]{Berry2007}
\bibinfo{author}{\bibfnamefont{D.}~\bibnamefont{Berry}},
  \bibinfo{author}{\bibfnamefont{G.}~\bibnamefont{Ahokas}},
  \bibinfo{author}{\bibfnamefont{R.}~\bibnamefont{Cleve}}, \bibnamefont{and}
  \bibinfo{author}{\bibfnamefont{B.}~\bibnamefont{Sanders}},
  \bibinfo{journal}{Commun. Math. Phys.} \textbf{\bibinfo{volume}{270}},
  \bibinfo{pages}{359} (\bibinfo{year}{2007}).

\bibitem[{\citenamefont{Childs}(2010)}]{Childs2010}
\bibinfo{author}{\bibfnamefont{A.}~\bibnamefont{Childs}},
  \bibinfo{journal}{Commun. Math. Phys.} \textbf{\bibinfo{volume}{294}},
  \bibinfo{pages}{581} (\bibinfo{year}{2010}).

\bibitem[{\citenamefont{Ambainis}(2010)}]{Ambainis2010}
\bibinfo{author}{\bibfnamefont{A.}~\bibnamefont{Ambainis}},
  \bibinfo{journal}{arXiv:1010.4458}  (\bibinfo{year}{2010}).

\bibitem[{\citenamefont{Kok et~al.}(2007)\citenamefont{Kok, Munro, Nemoto,
  Ralph, Dowling, and Milburn}}]{Kok2007}
\bibinfo{author}{\bibfnamefont{P.}~\bibnamefont{Kok}},
  \bibinfo{author}{\bibfnamefont{W.~J.} \bibnamefont{Munro}},
  \bibinfo{author}{\bibfnamefont{K.}~\bibnamefont{Nemoto}},
  \bibinfo{author}{\bibfnamefont{T.~C.} \bibnamefont{Ralph}},
  \bibinfo{author}{\bibfnamefont{J.~P.} \bibnamefont{Dowling}},
  \bibnamefont{and} \bibinfo{author}{\bibfnamefont{G.~J.}
  \bibnamefont{Milburn}}, \bibinfo{journal}{Rev. Mod. Phys.}
  \textbf{\bibinfo{volume}{79}}, \bibinfo{pages}{135} (\bibinfo{year}{2007}).

\bibitem[{\citenamefont{O'Brien and Akira~Furusawa}(2009)}]{OBrien2009}
\bibinfo{author}{\bibfnamefont{J.}~\bibnamefont{O'Brien}} \bibnamefont{and}
  \bibinfo{author}{\bibfnamefont{J.}~\bibnamefont{Akira~Furusawa},
  \bibfnamefont{Vu{\v{c}}kovi{\'c}}}, \bibinfo{journal}{Nature Photon.}
  \textbf{\bibinfo{volume}{3}}, \bibinfo{pages}{687} (\bibinfo{year}{2009}).

\bibitem[{\citenamefont{O'Brien et~al.}(2003)\citenamefont{O'Brien, Pryde,
  White, Ralph, and Branning}}]{OBrien2003}
\bibinfo{author}{\bibfnamefont{J.~L.} \bibnamefont{O'Brien}},
  \bibinfo{author}{\bibfnamefont{G.~J.} \bibnamefont{Pryde}},
  \bibinfo{author}{\bibfnamefont{A.~G.} \bibnamefont{White}},
  \bibinfo{author}{\bibfnamefont{T.~C.} \bibnamefont{Ralph}}, \bibnamefont{and}
  \bibinfo{author}{\bibfnamefont{D.}~\bibnamefont{Branning}},
  \bibinfo{journal}{Nature} \textbf{\bibinfo{volume}{426}},
  \bibinfo{pages}{264} (\bibinfo{year}{2003}).

\bibitem[{\citenamefont{Langford et~al.}(2005)\citenamefont{Langford, Weinhold,
  Prevedel, Resch, Gilchrist, O\char39{}Brien, Pryde, and
  White}}]{Langford2005}
\bibinfo{author}{\bibfnamefont{N.~K.} \bibnamefont{Langford}},
  \bibinfo{author}{\bibfnamefont{T.~J.} \bibnamefont{Weinhold}},
  \bibinfo{author}{\bibfnamefont{R.}~\bibnamefont{Prevedel}},
  \bibinfo{author}{\bibfnamefont{K.~J.} \bibnamefont{Resch}},
  \bibinfo{author}{\bibfnamefont{A.}~\bibnamefont{Gilchrist}},
  \bibinfo{author}{\bibfnamefont{J.~L.} \bibnamefont{O\char39{}Brien}},
  \bibinfo{author}{\bibfnamefont{G.~J.} \bibnamefont{Pryde}}, \bibnamefont{and}
  \bibinfo{author}{\bibfnamefont{A.~G.} \bibnamefont{White}},
  \bibinfo{journal}{Phys. Rev. Lett.} \textbf{\bibinfo{volume}{95}},
  \bibinfo{pages}{210504} (\bibinfo{year}{2005}).

\bibitem[{\citenamefont{Kiesel et~al.}(2005)\citenamefont{Kiesel, Schmid,
  Weber, Ursin, and Weinfurter}}]{Kiesel2005a}
\bibinfo{author}{\bibfnamefont{N.}~\bibnamefont{Kiesel}},
  \bibinfo{author}{\bibfnamefont{C.}~\bibnamefont{Schmid}},
  \bibinfo{author}{\bibfnamefont{U.}~\bibnamefont{Weber}},
  \bibinfo{author}{\bibfnamefont{R.}~\bibnamefont{Ursin}}, \bibnamefont{and}
  \bibinfo{author}{\bibfnamefont{H.}~\bibnamefont{Weinfurter}},
  \bibinfo{journal}{Phys. Rev. Lett.} \textbf{\bibinfo{volume}{95}},
  \bibinfo{pages}{210505} (\bibinfo{year}{2005}).

\bibitem[{\citenamefont{Okamoto et~al.}(2005)\citenamefont{Okamoto, Hofmann,
  Takeuchi, and Sasaki}}]{Okamoto2005}
\bibinfo{author}{\bibfnamefont{R.}~\bibnamefont{Okamoto}},
  \bibinfo{author}{\bibfnamefont{H.~F.} \bibnamefont{Hofmann}},
  \bibinfo{author}{\bibfnamefont{S.}~\bibnamefont{Takeuchi}}, \bibnamefont{and}
  \bibinfo{author}{\bibfnamefont{K.}~\bibnamefont{Sasaki}},
  \bibinfo{journal}{Phys. Rev. Lett.} \textbf{\bibinfo{volume}{95}},
  \bibinfo{pages}{210506} (\bibinfo{year}{2005}).

\bibitem[{\citenamefont{Zhou et~al.}(2011{\natexlab{a}})\citenamefont{Zhou,
  Ralph, Kalasuwan, Zhang, Peruzzo, Lanyon, and O'Brien}}]{Zhou2011a}
\bibinfo{author}{\bibfnamefont{X.}~\bibnamefont{Zhou}},
  \bibinfo{author}{\bibfnamefont{T.}~\bibnamefont{Ralph}},
  \bibinfo{author}{\bibfnamefont{P.}~\bibnamefont{Kalasuwan}},
  \bibinfo{author}{\bibfnamefont{M.}~\bibnamefont{Zhang}},
  \bibinfo{author}{\bibfnamefont{A.}~\bibnamefont{Peruzzo}},
  \bibinfo{author}{\bibfnamefont{B.}~\bibnamefont{Lanyon}}, \bibnamefont{and}
  \bibinfo{author}{\bibfnamefont{J.}~\bibnamefont{O'Brien}},
  \bibinfo{journal}{Nature Commun.} \textbf{\bibinfo{volume}{2}},
  \bibinfo{pages}{413} (\bibinfo{year}{2011}{\natexlab{a}}).

\bibitem[{\citenamefont{Crespi et~al.}(2011)\citenamefont{Crespi, Ramponi,
  Osellame, Sansoni, Bongioanni, Sciarrino, Vallone, and
  Mataloni}}]{Crespi2011}
\bibinfo{author}{\bibfnamefont{A.}~\bibnamefont{Crespi}},
  \bibinfo{author}{\bibfnamefont{R.}~\bibnamefont{Ramponi}},
  \bibinfo{author}{\bibfnamefont{R.}~\bibnamefont{Osellame}},
  \bibinfo{author}{\bibfnamefont{L.}~\bibnamefont{Sansoni}},
  \bibinfo{author}{\bibfnamefont{I.}~\bibnamefont{Bongioanni}},
  \bibinfo{author}{\bibfnamefont{F.}~\bibnamefont{Sciarrino}},
  \bibinfo{author}{\bibfnamefont{G.}~\bibnamefont{Vallone}}, \bibnamefont{and}
  \bibinfo{author}{\bibfnamefont{P.}~\bibnamefont{Mataloni}},
  \bibinfo{journal}{Nature Commun.} \textbf{\bibinfo{volume}{2}},
  \bibinfo{pages}{566} (\bibinfo{year}{2011}).

\bibitem[{\citenamefont{Sansoni et~al.}(2010)\citenamefont{Sansoni, Sciarrino,
  Vallone, Mataloni, Crespi, Ramponi, and Osellame}}]{Sansoni2010}
\bibinfo{author}{\bibfnamefont{L.}~\bibnamefont{Sansoni}},
  \bibinfo{author}{\bibfnamefont{F.}~\bibnamefont{Sciarrino}},
  \bibinfo{author}{\bibfnamefont{G.}~\bibnamefont{Vallone}},
  \bibinfo{author}{\bibfnamefont{P.}~\bibnamefont{Mataloni}},
  \bibinfo{author}{\bibfnamefont{A.}~\bibnamefont{Crespi}},
  \bibinfo{author}{\bibfnamefont{R.}~\bibnamefont{Ramponi}}, \bibnamefont{and}
  \bibinfo{author}{\bibfnamefont{R.}~\bibnamefont{Osellame}},
  \bibinfo{journal}{Phys. Rev. Lett.} \textbf{\bibinfo{volume}{105}},
  \bibinfo{pages}{200503} (\bibinfo{year}{2010}).

\bibitem[{\citenamefont{Pittman et~al.}(2001)\citenamefont{Pittman, Jacobs, and
  Franson}}]{Pittman2001}
\bibinfo{author}{\bibfnamefont{T.}~\bibnamefont{Pittman}},
  \bibinfo{author}{\bibfnamefont{B.}~\bibnamefont{Jacobs}}, \bibnamefont{and}
  \bibinfo{author}{\bibfnamefont{J.}~\bibnamefont{Franson}},
  \bibinfo{journal}{Phys.\ Rev.\ A} \textbf{\bibinfo{volume}{64}},
  \bibinfo{pages}{062311} (\bibinfo{year}{2001}).

\bibitem[{\citenamefont{Gasparoni et~al.}(2004)\citenamefont{Gasparoni, Pan,
  Walther, Rudolph, and Zeilinger}}]{Gasparoni2004}
\bibinfo{author}{\bibfnamefont{S.}~\bibnamefont{Gasparoni}},
  \bibinfo{author}{\bibfnamefont{J.-W.} \bibnamefont{Pan}},
  \bibinfo{author}{\bibfnamefont{P.}~\bibnamefont{Walther}},
  \bibinfo{author}{\bibfnamefont{T.}~\bibnamefont{Rudolph}}, \bibnamefont{and}
  \bibinfo{author}{\bibfnamefont{A.}~\bibnamefont{Zeilinger}},
  \bibinfo{journal}{Phys. Rev. Lett.} \textbf{\bibinfo{volume}{93}},
  \bibinfo{pages}{020504} (\bibinfo{year}{2004}).

\bibitem[{\citenamefont{Lanyon et~al.}(2010)\citenamefont{Lanyon, Whitfield,
  Gillett, Goggin, Almeida, Kassal, Biamonte, Mohseni, Powell, Barbieri
  et~al.}}]{Lanyon2010}
\bibinfo{author}{\bibfnamefont{B.}~\bibnamefont{Lanyon}},
  \bibinfo{author}{\bibfnamefont{J.}~\bibnamefont{Whitfield}},
  \bibinfo{author}{\bibfnamefont{G.}~\bibnamefont{Gillett}},
  \bibinfo{author}{\bibfnamefont{M.}~\bibnamefont{Goggin}},
  \bibinfo{author}{\bibfnamefont{M.}~\bibnamefont{Almeida}},
  \bibinfo{author}{\bibfnamefont{I.}~\bibnamefont{Kassal}},
  \bibinfo{author}{\bibfnamefont{J.}~\bibnamefont{Biamonte}},
  \bibinfo{author}{\bibfnamefont{M.}~\bibnamefont{Mohseni}},
  \bibinfo{author}{\bibfnamefont{B.}~\bibnamefont{Powell}},
  \bibinfo{author}{\bibfnamefont{M.}~\bibnamefont{Barbieri}},
  \bibnamefont{et~al.}, \bibinfo{journal}{Nature Chem.}
  \textbf{\bibinfo{volume}{2}}, \bibinfo{pages}{106} (\bibinfo{year}{2010}).

\bibitem[{\citenamefont{Zhou et~al.}(2011{\natexlab{b}})\citenamefont{Zhou,
  Kalasuwan, Ralph, and O'Brien}}]{Zhou2011}
\bibinfo{author}{\bibfnamefont{X.}~\bibnamefont{Zhou}},
  \bibinfo{author}{\bibfnamefont{P.}~\bibnamefont{Kalasuwan}},
  \bibinfo{author}{\bibfnamefont{T.}~\bibnamefont{Ralph}}, \bibnamefont{and}
  \bibinfo{author}{\bibfnamefont{J.}~\bibnamefont{O'Brien}},
  \bibinfo{journal}{arXiv:1110.4276}  (\bibinfo{year}{2011}{\natexlab{b}}).

\bibitem[{\citenamefont{Mart{\'\i}n-L{\'o}pez
  et~al.}(2012)\citenamefont{Mart{\'\i}n-L{\'o}pez, Laing, Lawson, Alvarez,
  Zhou, and O'Brien}}]{Martin-Lopez2012}
\bibinfo{author}{\bibfnamefont{E.}~\bibnamefont{Mart{\'\i}n-L{\'o}pez}},
  \bibinfo{author}{\bibfnamefont{A.}~\bibnamefont{Laing}},
  \bibinfo{author}{\bibfnamefont{T.}~\bibnamefont{Lawson}},
  \bibinfo{author}{\bibfnamefont{R.}~\bibnamefont{Alvarez}},
  \bibinfo{author}{\bibfnamefont{X.}~\bibnamefont{Zhou}}, \bibnamefont{and}
  \bibinfo{author}{\bibfnamefont{J.}~\bibnamefont{O'Brien}},
  \bibinfo{journal}{Nature Photon.}  (\bibinfo{year}{2012}).

\bibitem[{\citenamefont{Leyton and Osborne}(2008)}]{Leyton2008}
\bibinfo{author}{\bibfnamefont{S.~K.} \bibnamefont{Leyton}} \bibnamefont{and}
  \bibinfo{author}{\bibfnamefont{T.~J.} \bibnamefont{Osborne}},
  \bibinfo{journal}{arXiv:0812.4423}  (\bibinfo{year}{2008}).

\bibitem[{\citenamefont{Wiebe et~al.}(2012)\citenamefont{Wiebe, Braun, and
  Lloyd}}]{Wiebe2012}
\bibinfo{author}{\bibfnamefont{N.}~\bibnamefont{Wiebe}},
  \bibinfo{author}{\bibfnamefont{D.}~\bibnamefont{Braun}}, \bibnamefont{and}
  \bibinfo{author}{\bibfnamefont{S.}~\bibnamefont{Lloyd}},
  \bibinfo{journal}{Phys. Rev. Lett.} \textbf{\bibinfo{volume}{109}},
  \bibinfo{pages}{050505} (\bibinfo{year}{2012}).

\bibitem[{\citenamefont{Lanyon et~al.}(2009)\citenamefont{Lanyon, Barbieri,
  Almeida, Jennewein, Ralph, Resch, Pryde, O'Brien, Gilchrist, and
  White}}]{Lanyon2009}
\bibinfo{author}{\bibfnamefont{B.~P.} \bibnamefont{Lanyon}},
  \bibinfo{author}{\bibfnamefont{M.}~\bibnamefont{Barbieri}},
  \bibinfo{author}{\bibfnamefont{M.~P.} \bibnamefont{Almeida}},
  \bibinfo{author}{\bibfnamefont{T.}~\bibnamefont{Jennewein}},
  \bibinfo{author}{\bibfnamefont{T.~C.} \bibnamefont{Ralph}},
  \bibinfo{author}{\bibfnamefont{K.~J.} \bibnamefont{Resch}},
  \bibinfo{author}{\bibfnamefont{G.~J.} \bibnamefont{Pryde}},
  \bibinfo{author}{\bibfnamefont{J.~L.} \bibnamefont{O'Brien}},
  \bibinfo{author}{\bibfnamefont{A.}~\bibnamefont{Gilchrist}},
  \bibnamefont{and} \bibinfo{author}{\bibfnamefont{A.~G.} \bibnamefont{White}},
  \bibinfo{journal}{Nature Phys.} \textbf{\bibinfo{volume}{5}},
  \bibinfo{pages}{134} (\bibinfo{year}{2009}).

\bibitem[{\citenamefont{Kwiat et~al.}(1999)\citenamefont{Kwiat, Waks, White,
  Appelbaum, and Eberhard}}]{Kwiat1999}
\bibinfo{author}{\bibfnamefont{P.~G.} \bibnamefont{Kwiat}},
  \bibinfo{author}{\bibfnamefont{E.}~\bibnamefont{Waks}},
  \bibinfo{author}{\bibfnamefont{A.~G.} \bibnamefont{White}},
  \bibinfo{author}{\bibfnamefont{I.}~\bibnamefont{Appelbaum}},
  \bibnamefont{and} \bibinfo{author}{\bibfnamefont{P.~H.}
  \bibnamefont{Eberhard}}, \bibinfo{journal}{Phys. Rev. A}
  \textbf{\bibinfo{volume}{60}}, \bibinfo{pages}{R773} (\bibinfo{year}{1999}).

\end{thebibliography}

{\bf Acknowledgments}
We thank Frank Verstraete for valuable discussions and Thomas Lindner for assistance in the laboratory.
I.K. was supported by a UQ Postdoctoral Research Fellowship and acknowledges support from the Australian Research Council Centres of Excellence for Engineered Quantum Systems (project CE110001013) and Quantum Computation and Communication Technology (project CE110001027).
A.A.G. acknowledges support from the Hughes Research Laboratory (grant M1144-201167-DS) and the Air Force Research Office (contract FA9550-12-1-0046
and 10323836-SUB) and thanks the Corning, Sloan and Dreyfus foundations for their support.
P.W. acknowledges support from the European Commission, Q-ESSENCE (No. 248095), QUILMI (No. 295293) and the ERA-Net CHISTERA project QUASAR, the John Templeton Foundation, the Vienna Center for Quantum Science and Technology (VCQ), the Austrian Nano-initiative NAP Platon, the Austrian Science Fund (FWF) through the SFB FoQuS (No. F4006-N16), START (No. Y585- N20) and the doctoral programme CoQuS, the Vienna Science and Technology Fund (WWTF) under grant ICT12-041, and the Air Force Office of Scientific Research, Air Force Material Command, United States Air Force, under grant number FA8655-11-1-3004.
\\ \\

%%%%%%%%%%%%%%%%%%%%%%%%%%%%%%%%%%%%%%%%%%%%%%%%%%%%%%%%%%%%%%%%%%%%%%%%%%%%%%%%%%%%%%%%%%%%%%%%%%%%%%%%%%%%%%%%%%%%%%%%%%%%%%%%%%%%%%%%%%%%%%%%%%%%%%
\section{Appendix}
%%%%%%%%%%%%%%%%%%%%%%%%%%%%%%%%%%%%%%%%%%%%%%%%%%%%%%%%%%%%%%%%%%%%%%%%%%%%%%%%%%%%%%%%%%%%%%%%%%%%%%%%%%%%%%%%%%%%%%%%%%%%%%%%%%%%%%%%%%%%%%%%%%%%%%

\subsection{Theory}

Our work differs from the original proposal in~\cite{Harrow2009} in several modifications that are needed to 
implement the algorithm with a limited number of qubits. Some of these modifications were mentioned in the main text, and we
elaborate on them here, essentially describing the simplifications that are used to transform the circuit in Figure~1a to the circuit in
Figure~1c.

As described in the text, the most general circuit involves phase estimation, a controlled $R_y$ rotation, and a reverse phase estimation
(for a description of phase estimation in general, see Sec.~5.2 of~\cite{Nielsen2000}). If we restrict ourselves to one state qubit and one
eigenvalue qubit, phase estimation is simply a controlled unitary between two Hadamard gates on the eigenvalue qubit (see Figure~1a).

The unitary $U$ that is chosen depends on what binary digit of the eigenvalue needs to be read out. Because we have assumed
that $A$ is scaled in such a way that its eigenvalues lie in the range $(0,1)$, we can express them as follows:
\begin{align}
\lambda_1 &= 0.\overline{x_{1,1} x_{1,2} x_{1,3} x_{1,4}} \ldots \\
\lambda_2 &= 0.\overline{x_{2,1} x_{2,2} x_{2,3} x_{2,4}} \ldots ,
\end{align}
where the $\bar{x}_{i,j}$ are binary digits, 0 or 1. If we wish to read out the $n^\mathrm{th}$ digit, 
we must choose $U=\exp\left(2\pi i \,2^{n-1} A\right)$. Note its action on the eigenvectors:
\begin{align}
U\ket{u_j}&=\exp\left(2\pi i \,2^{n-1} \lambda_j\right)\ket{u_j}\\
&=\exp\left(2\pi i \,2^{n-1} 0.\overline{x_{j,1} x_{j,2} x_{j,3} x_{j,4}} \ldots\right)\ket{u_j}\\
&=\exp\left(2\pi i \, 0.\overline{x_{j,n} x_{j,n+1} x_{j,n+2}} \ldots\right)\ket{u_j}\\
&\approx \exp\left(2\pi i \, 0.\bar{x}_{j, n} \right)\ket{u_j}\\
&=(-1)^{\bar{x}_{j,n}} \ket{u_j}.
\end{align}
The approximation of neglecting all digits past $\bar{x}_n$ introduces errors to the procedure,
which would be mitigated by using additional eigenvalue qubits. Without the additional qubits, we avoid the error
by assuming, as in the main text, that the eigenvalues are of the form $0.\bar{a}0$ and $0.\bar{a}1$,
differing only at the $n^\mathrm{th}$ digit.

The state and eigenvalue qubits, initialized to $\ket{b}_S\ket{0}_E$, transform thus:
\begin{align}
\ket{b}_S\ket{0}_E &= \sum_{j=1}^2 \beta_j \ket{u_j}\ket{0}\xrightarrow{H} \sum_{j=1}^2 \beta_j \ket{u_j}\frac{\ket{0}+\ket{1}}{\sqrt{2}}\\
&\xrightarrow{\mathrm{cont.-}U} \sum_{j=1}^2 \beta_j \ket{u_j}\frac{\ket{0}+(-1)^{\bar{x}_{j,n}}\ket{1}}{\sqrt{2}} \\
&\xrightarrow{H} \sum_{j=1}^2 \beta_j \ket{u_j}\ket{\bar{x}_{j,n}}.
\end{align}
The two qubits are now entangled, with each eigenstate in the state register accompanied by the
$n^\mathrm{th}$ digit of its eigenvalue.

As described in the main text, $\theta=-2\arccos(\lambda_1/\lambda_2)$ in all cases. For example, if, as in Figure 3, 
$\lambda_1=0.10=\frac12$ and $\lambda_2=0.11=\frac34$, in which case we read out the second digit, we get $\theta=-1.682$. 
In the second column of Figure 4, $\lambda_1=0.100=\frac12$ and $\lambda_2=0.101=\frac58$, and so $\theta=-1.287$
(with the third digit read out). In the third column of Figure 4, $\lambda_1=0.110=\frac12$ and $\lambda_2=0.111=\frac58$, and so $\theta=-1.082$.

%------------------------------------------------------------------------------------------------------------------------------
\begin{figure*}%[th]
\centering
\includegraphics[width = .95\columnwidth]{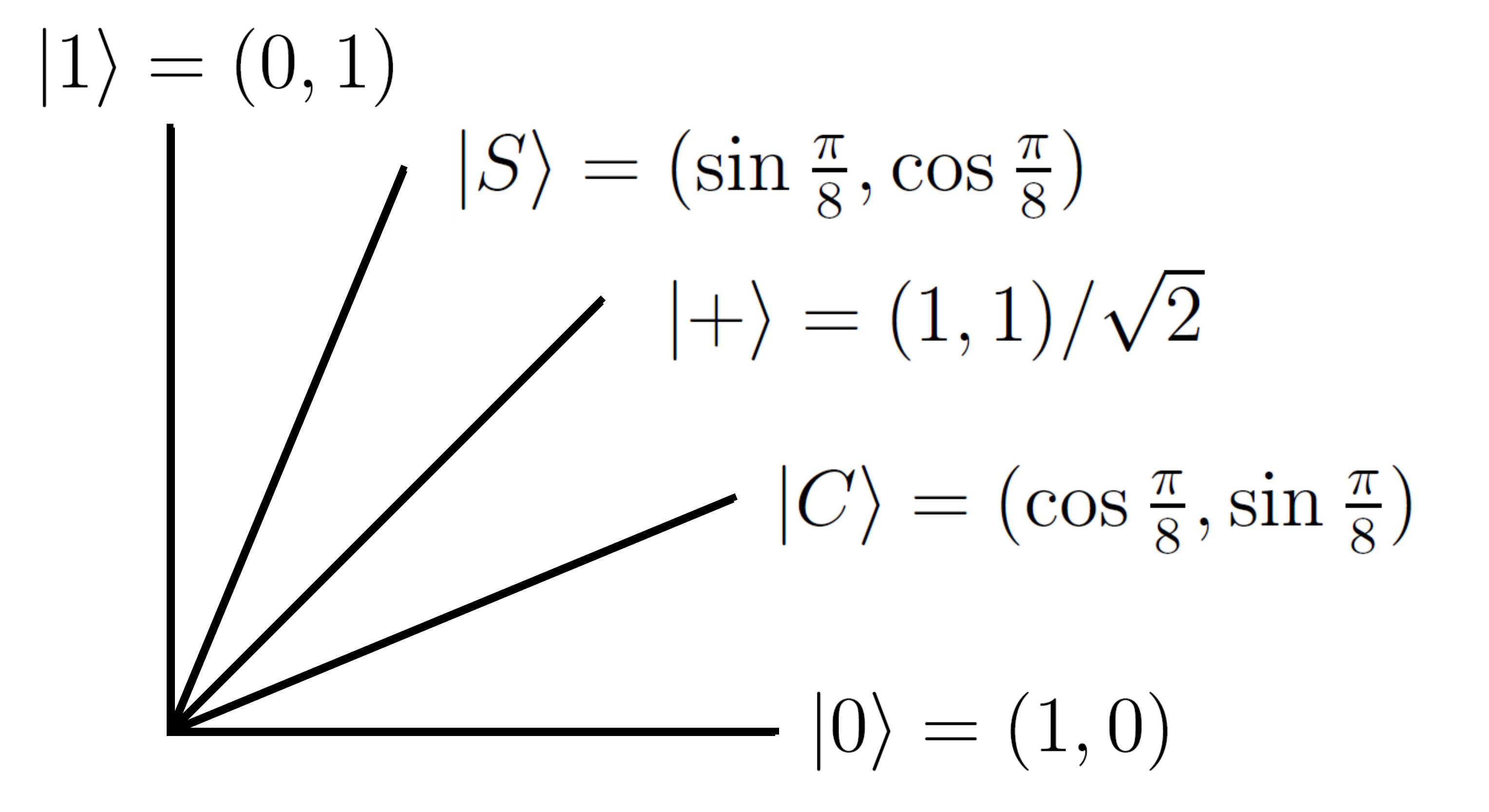}
\caption[~]{The figure shows the different states $R\ket{b}$ which we have chosen as input states to our circuit.
\label{SIFigure1}}
\end{figure*}
%------------------------------------------------------------------------------------------------------------------------------

Once we assume that the eigenvalues differ in the $n^\mathrm{th}$ digit, the action of $U$ is seen to be very simple: it is merely a 
$Z$ gate in the eigenbasis of $A$. That is, it leaves the first eigenstate unchanged and adds a phase of $-1$ to the second. It
can therefore be written as $U=R^\dagger ZR$, where $R$ diagonalizes $A$. It is with this fact, and remembering that a
controlled-$Z$ conjugated with Hadamard gates gives a controlled-NOT, that one obtains the circuit in Figure~1b.

To get from Figure~1b to Figure~1c, we first observe that the eigenvalue qubit can be removed. After the gate $R$, 
the first CNOT transfers the state of the state register to the eigenvalue register. This then operates the controlled-$R_y(\theta)$ rotation. 
Instead of transferring the state first, it is equivalent to simply control the $R_y(\theta)$ gate directly from the state qubit, giving this circuit:
\[
\Qcircuit @C=0.4em @R=0.3em @!R 
{\lstick{|b\rangle} & \gate{R} & \ctrl{1} & \gate{R^\dagger} & \qw & \rstick{|x\rangle} \qw \\
\lstick{|1\rangle} & \qw & \gate{R_y(\theta)}  & \qw & \meter & \rstick{1} \cw  }
\]
Controlled single-qubit rotations have already been implemented in linear optics~\cite{Lanyon2009, Lanyon2010, Zhou2011}, but here we follow a different approach.
We decompose the controlled-$R_y$ rotation using the general method described in Sec.~4.3 of~\cite{Nielsen2000}, which immediately
gives the circuit in Figure 1c. 
%%%%%%%%%%%%%%%%%%%%%%%%%%%%%%%%%%%%%%%%%%%%%%%%%%%%%%%%%%%%%%%%%%%%%%%%%%%%%%%%%%%%%%%%%%%%%%%%%%%%%%%%%%%%%%%%%%%%%%%%%%%%%%%%%%%%%%%%%%%%%%%%%%%%%%
\subsection{Experimental setup}
In our experiment, entangled photon pairs are produced by exploiting
the emissions of a non-collinear type-II SPDC process~\cite{Kwiat1999}. For this, a mode-locked Mira HP Ti:Sa
oscillator is pumped by a Coherent Inc. Verdi V-10 laser. The pulsed-laser output ($\tau=200\,$fs, $\lambda=789\,$nm, 76 MHz) is frequency-doubled using a 2~mm-thick lithium triborate (LBO) crystal,
resulting in UV pulses of 0.75~W cw average. We achieve a stable source of UV pulses by translating
the LBO to avoid optical damage to the anti-reflection coating of the crystal.
The UV laser beam passes through a 2~mm-thick $\beta$-barium borate crystal, gets reflected, and passes through the crystal a second time (see Figure~\ref{Figure2}). The photons created during the first pass of the laser beam enter the first CNOT gate as the input (control and target) qubits. The state of these input qubits is modified using polarisers and additional local unitary gates that in principle allow for the creation of arbitrary input states. 
In our experiment, the control and target qubits are prepared in the states $R|b\rangle$ and $|1\rangle$, respectively.
Thus, we absorb the local operation $R$ in the preparation of the input state.
The photons created when the laser passes through the crystal the second time act as the entangled ancilla photon pairs which is required for the first CNOT gate. For this, we align our setup such that the entangled state  $\ket{\Phi^+} = (\ket{00}+\ket{11})/\sqrt{2}$ is emitted~\cite{Gasparoni2004}. 

The photons interfere at the polarising beam splitters (PBS) as shown in Figure~\ref{Figure4}.
The PBS on the control side (target side) is aligned such that it acts in the basis $\left\{\ket{0},\ket{1}\right\}$ ($\left\{\ket{+},\ket{-}\right\}$). The photons are the filtered spatially and spectrally with the help of narrow-band filters ($\Delta\lambda=3\,$nm) and by coupling them into single-mode fibers. A coincidence detection of the ancilla qubits in detectors $3$ and $4$ in the state $\ket{-}_3\ket{1}_4$ signals a successful gate operation. 

The output photons in the modes  I and II (see Figure~\ref{Figure4}) are then guided to the second CNOT gate.
Wave plates before and after the second CNOT implement the local rotations $R_y(\theta/2)$, $R_y(\theta/2)$, $R$, and $R^\dagger$. 

Photons are coupled out, pass the polarisation-dependent beam splitters (PDBSs), and are coupled to single-mode fibers again.
The success of the second CNOT operation is determined by postselection on a coincidence detection in outputs $1$ and $2$.
The algorithm succeeds if the target photon is detected in state $|1\rangle$. The output control qubit is in the state $|x\rangle$, which is analysed by using HWPs, QWPs and polarising beam splitters; and a full state tomography of the output state $\ket{x}$ is performed.
Errors are obtained from a Monte Carlo routine assuming Poissonian counting statistics. These indicate a lower bound for the actual error that takes
all the experimental imperfections into account.
In our experiment, typical visibilities of the emitted Bell pairs are about 0.9 and higher-order emissions degrade the quality of our gate operations.
The back-reflecting mirror is continuously moved back-and-forth to avoid any phase correlations between of the signal and the noise originating from higher-order photon emissions.
Additionally, imperfect visibilities on the order of 0.85 to 0.9 of the quantum interference at the PBSs in the first gate and the PDBS in the second gate contribute to errors. 
%\todo[inline]{mention countrates}

%%%%%%%%%%%%%%%%%%%%%%%%%%%%%%%%%%%%%%%%%%%%%%%%%%%%%%%%%%%%%%%%%%%%%%%%%%%%%%%%%%%%%%%%%%%%%%%%%%%%%%%%%%%%%%%%%%%%%%%%%%%%%%%%%%%%%%%%%%%%%%%%%%%%%%

\subsection{Experiment}
 
\begin{figure*}[th]
\centering
\includegraphics[width = 1.6\columnwidth]{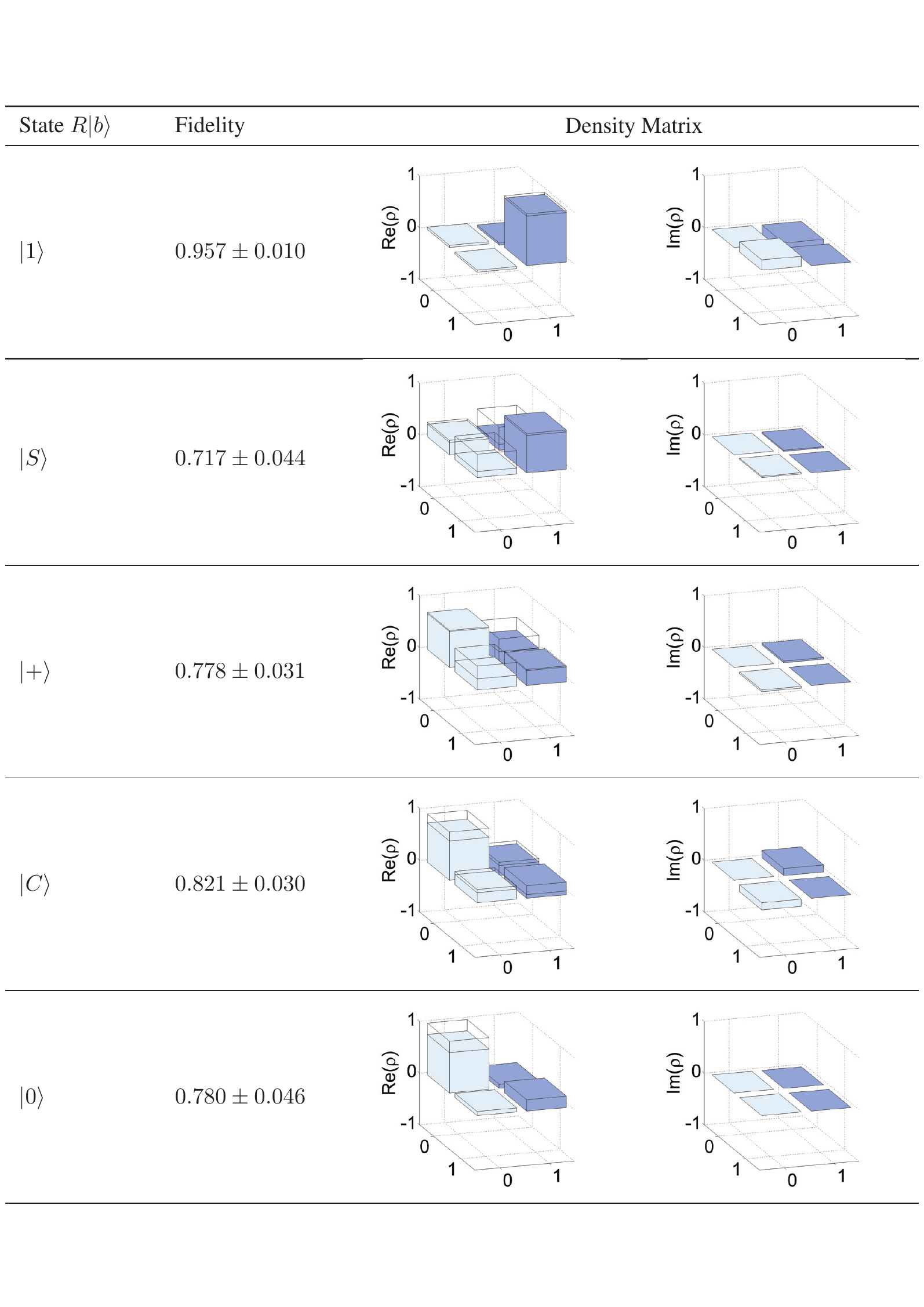}
\caption[~]{The figure shows the different states $R\ket{b}$ which we have chosen as input states to our circuit.
\label{SITable1}}
\end{figure*}

\begin{figure*}[th]
\centering
\includegraphics[width = 1.6\columnwidth]{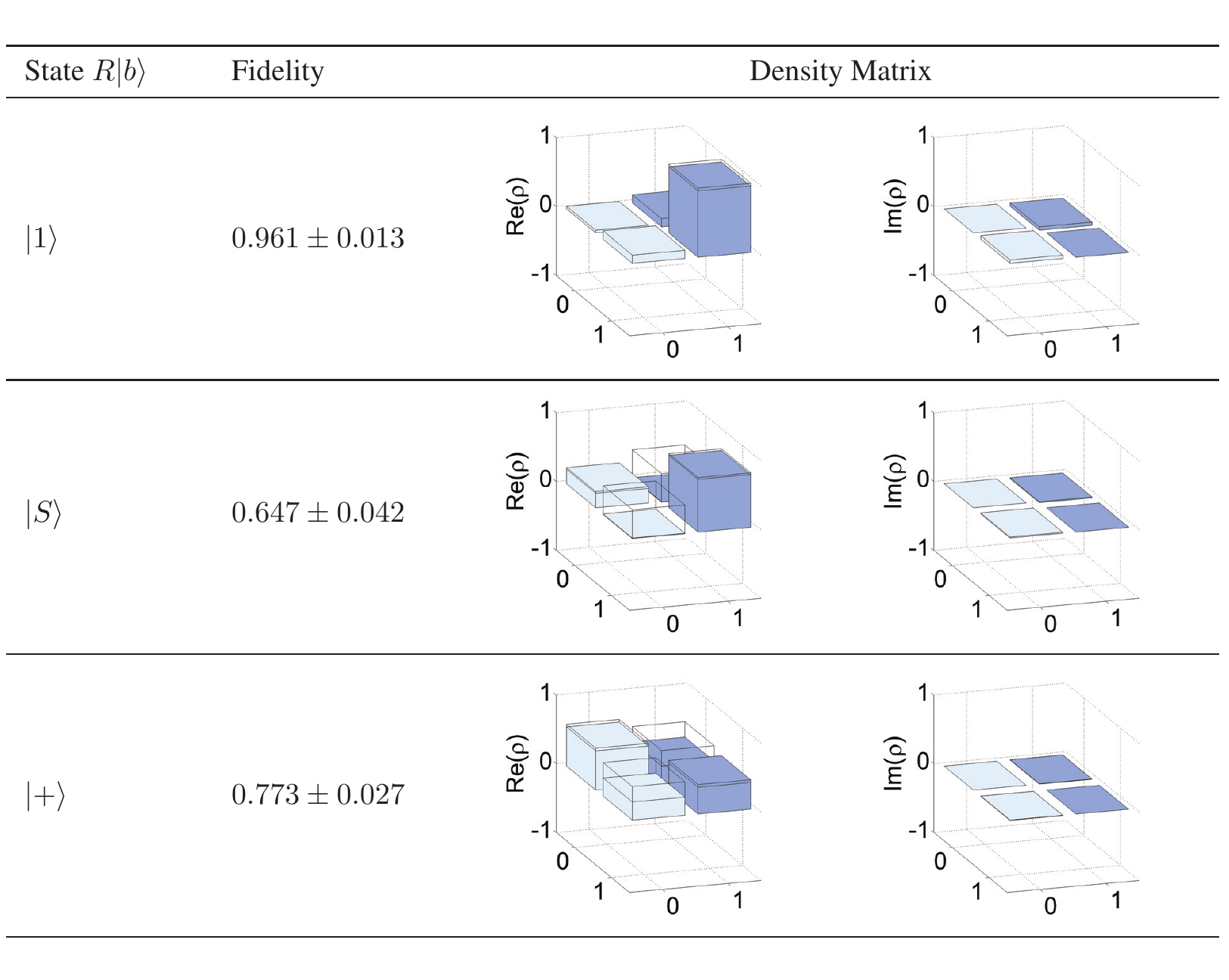}
\caption[~]{The figure shows the different states $R\ket{b}$ which we have chosen as input states to our circuit.
\label{SITable2}}
\end{figure*}

\begin{figure*}[th]
\centering
\includegraphics[width = 1.6\columnwidth]{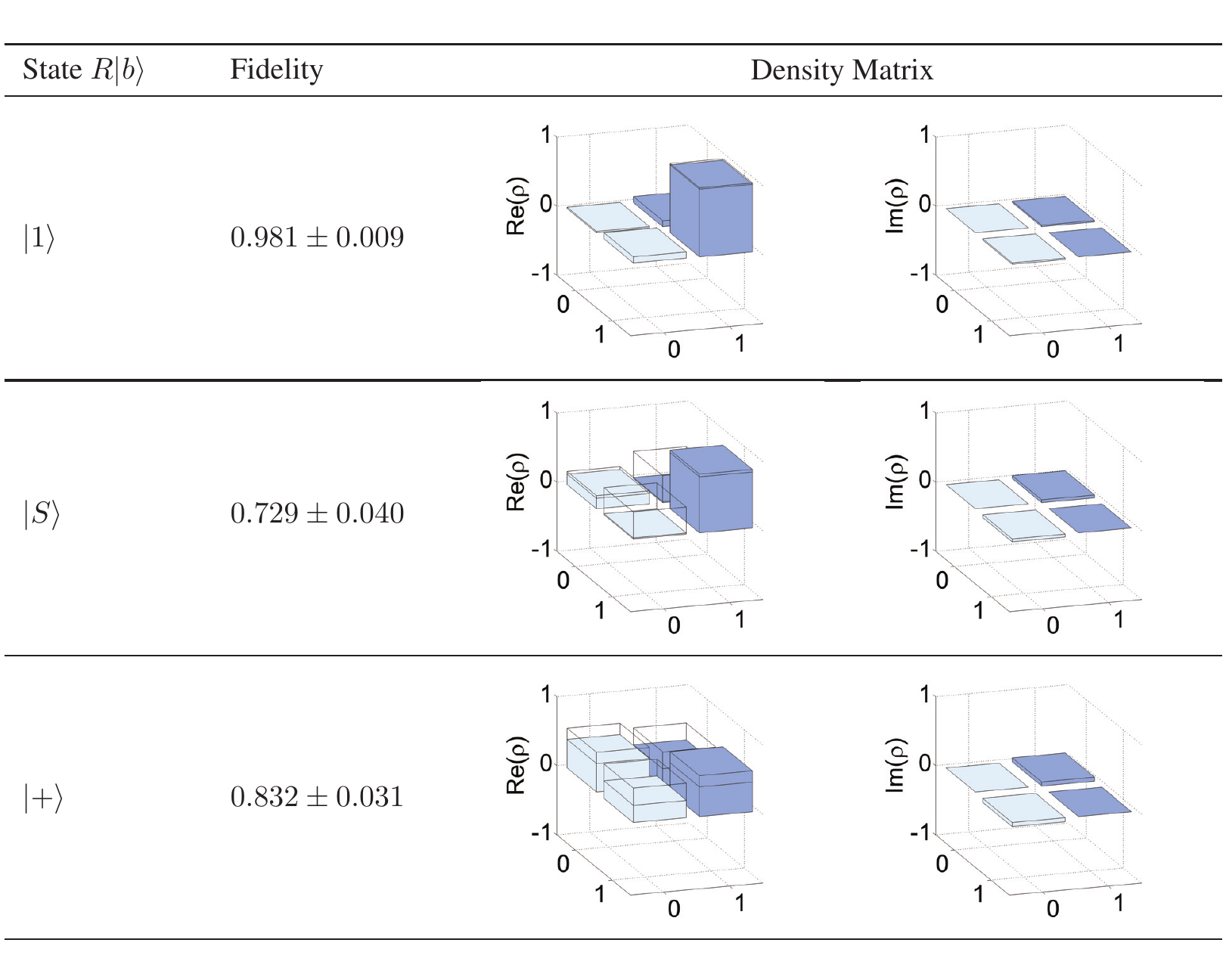}
\caption[~]{The figure shows the different states $R\ket{b}$ which we have chosen as input states to our circuit.
\label{SITable3}}
\end{figure*}

As mentioned in the main text, the fidelity of the output state $\ket{x}$ depends on the state $R\ket{b}$ that effectively enters the control input of the first CNOT gate. The Figures~\ref{SITable1},~\ref{SITable2}, and~\ref{SITable3} show the characterization of the state $\ket{x}$ for various input states. Figure~\ref{SIFigure1} depicts the different input states we have chosen for our analysis.

Our analysis shows that the fidelities of the obtained output states vary from $(64.7\pm4.2)\%$ to $(98.1\pm0.9)\%$.
These variations in fidelity arise due to the influence of higher-order emissions from spontaneous parametric down-conversion. These higher-order emissions can either occur when the beam passes the crystal the first time (``double-forward emission'') or when the beam passes the crystal the second time (``double-backward emission'').

If the input to the first CNOT gate is chosen to be $R\ket{b}=\ket{0}$ or $R\ket{b}=\ket{1}$, a double-forward emission can never lead to a fourfold coincidence and thus signal a wrong ``successful'' operation of the first gate. In these cases, the photons entering the control input to the gate will either both be reflected or both be transmitted at the first PBS. For all other inputs to the gate, these higher-order emissions degrade the fidelity of the output state as demonstrated in our analysis.

The ``double-backward emission'' can in principle never lead to a fourfold coincidence because the photons can never split up due to quantum interference. However, due to the visibility of the entangled Bell pairs of 0.9, this quantum interference does not work perfectly. About $10\%$ of our total counts arise from these events, which also influences the fidelity of the output state $\ket{x}$.
However, for the input $R\ket{b}=\ket{1}$ the fidelity does not seem to be affected. In this case, the noise cannot be distinguished from the signal.
The influence of the double-backward emission increases for the other input states and reaches a maximum for $R\ket{b}=\ket{0}$.

\end{document}